\documentstyle[aps,prb]{revtex}

\begin{document}

\setlength{\topmargin}{0mm}

\input epsf.sty
\twocolumn[\hsize\textwidth\columnwidth\hsize\csname %
@twocolumnfalse\endcsname

\draft

\widetext
%%%%%%%% prl (above) %%%%%%%%%%%%%%%%%%

\title{Systematic Study of Short Range Antiferromagnetic Order and 
The Spin-Glass State in Lightly Doped La$_{2-x}$Sr$_{x}$CuO$_{4}$}

\author{S. Wakimoto,\footnote{Present address: Massachusetts
Institute of Technology, Cambridge, MA 02139, USA} S. Ueki,
and Y. Endoh}
\address{Department of Physics, Tohoku University, Sendai 980-8578, Japan}
\author{K. Yamada}
\address{Institute for Chemical Research, Kyoto University, Gokasho, Uji 610-0011, Japan}

\date{\today}
\maketitle

\vspace{-0.1in}

\begin{abstract}

Systematic measurements of the magnetic susceptibility were performed on 
single crystals of lightly doped La$_{2-x}$Sr$_{x}$CuO$_{4}$ 
($x=0.03, 0.04$ and $0.05$).
For all samples the temperature dependence of the in-plane magnetic susceptibility 
shows typical spin-glass features with spin-glass transition temperatures $T_{g}$ 
of 6.3~K, 5.5~K and 5.0~K for $x=0.03, 0.04$ and $0.05$, respectively.
The canonical spin-glass order parameter extracted from the 
in-plane susceptibility of all the samples follows a universal scaling curve. 
On the other hand, 
the out-of-plane magnetic susceptibility deviates from Curie law
below a temperature $T_{dv}$, higher than $T_{g}$.
Comparing with previous neutron scattering results with an instrumental energy 
resolution of $\Delta\omega \sim 0.25$~meV from Wakimoto {\it et al.},
the $x$-dependence of $T_{dv}$ is qualitatively the same as that of 
$T_{el}$($\Delta\omega = 0.25$~meV), the temperature below which 
the elastic magnetic scattering develops around $(\pi, \pi)$.
Thus, a revised magnetic phase diagram in the lightly doped 
region of La$_{2-x}$Sr$_{x}$CuO$_{4}$ is proposed.
The Curie constants calculated from the in-plane
susceptibility are independent of the Sr concentration.
On the basis of the cluster spin-glass model, this fact might reflect an 
inhomogeneous distribution of doped holes in the CuO$_{2}$ plane, such as 
in a stripe structure.

\end{abstract}

\pacs{74.72.Dn, 74.25.Ha, 75.50.Lk}

%%%%% prl format (below) %%%%%%%%%%%%%%
\phantom{.}
]
\narrowtext
%%%%%%%% prl (above) %%%%%%%%%%%%%%%%%%

%
% Wakimoto 3-5% static results paper
%
% Edited by S. Wakimoto on Sep. 10, 1998 at Tohoku
% Edited by Y. Endoh on Sep. 15, 1998 at Tokyo
% Edited by S. Wakimoto on April 2, 1999 at MIT
% Edited by S. Wakimoto on April 19, 1999 at MIT
% Edited by S. Wakimoto on April 26, 1999 at NIST
% Edited by S. Wakimoto on July 14, 1999 at BNL
% Edited by Y. Endoh on July 15, 1999 at BNL
% Edited by S. Wakimoto on Aug. 18, 1999 at MIT
% Edited by S. Wakimoto on Aug. 28, 1999 at MIT
% Edited by R. Leheny and S. Wakimoto on Aug. 30 - Sept. 7, 1999 at MIT
% Edited by S. Wakimoto on Sept. 10, 1999 at MIT
%

\section{Introduction}

The single-layered high-temperature superconducting (hereafter 
abbreviated as HTSC) 2-1-4 type cuprates have 
received intensive attention in explorations not only of 
the microscopic HTSC mechanism but also of the basic properties of 
two dimensional magnetism in the square-lattice Heisenberg 
antiferromagnet.~\cite{M.A.Kastner_98} 
The three-dimensional (3D) antiferromagnetic (AF) long range 
order in undoped La$_{2}$CuO$_{4}$ quickly vanishes~\cite{Aharony_88} 
with hole doping either through the substitution of 
La$^{3+}$ sites with Sr$^{2+}$ or by the insertion of the excess oxygen.
Upon further doping, the La$_{2-x}$Sr$_{x}$CuO$_{4}$ (LSCO) system shows superconductivity 
in the range $0.06 \stackrel{<}{\sim} x \stackrel{<}{\sim} 0.25$ 
as well as incommensurate spin fluctuations.~\cite{K.Yamada_98,sum}
Static spin correlations with the same incommensurability as those of dynamic one 
are also observed for the samples whose doping rates are  
near $x=0.12$~\cite{T.Suzuki_98,H.Kimura_98} and $x=0.06$.~\cite{waki0.06}
In the intermediate regime, $0.02 \leq x \leq 0.05$, placed between the
3D AF state and the superconducting state, there coexists a canonical 
spin-glass state observed by magnetic susceptibility 
measurements~\cite{F.C.Chou_95} and quasi-static magnetic order 
investigated by neutron scattering experiments.~\cite{Hayden_91,B.Keimer_92,waki_neut}
%*******ADDITIONAL PART**********
From the $^{139}$La NQR measurements, 
spin glass state with magnetically ordered finite-size regions, i.e. 
cluster spin glass state, is argued.~\cite{Cho_92PRB}
It is also reported that spin glass freezing temperature determined by 
the $^{139}$La NQR measurements in this region is proportional to 
$1/x$.~\cite{Chou_93PRL}
%*********************************
Very recently, Wakimoto {\it et al.}~\cite{waki_full} found that the quasi-static 
magnetic order for $x=0.05$ has an incommensurate structure whose modulation 
vector is along the orthorhombic $b$-axis, $\sim 45^{\circ}$ away from the incommensurate 
modulation observed in superconducting samples.
Furthermore, an extensive study by Matsuda {\it et al.}~\cite{Matsuda_incomme} 
has revealed that the same type of incommensurate spin structure exists 
throughout the hole concentration range $0.024 \leq x \leq 0.05$.

From numerical simulations Gooding {\it et al.}~\cite{Gooding} explained 
the coexistence of the spin-glass state and quasi-static magnetic order 
as a cluster spin-glass state taking into account hole-localization around Sr ions.
However, important unresolved problems exist for the physics in this intermediate 
doping region. 
One issue is how the incommensurate magnetic state is realized in the spin-glass state.
For example, the model proposed by Gooding {\it et al.} predicts commensurate 
spin correlations in this region.
Therefore, a new model for the spin state seems needed.
Another important question is how the spin-glass state changes from the insulating region
$(x \leq 0.05)$ to superconducting region $(x \geq 0.06)$.
In spite of the dramatic change of the quasi-static correlations 
at the insulator-superconductor boundary from a diagonal 
incommensurate state, which is modulated along the diagonal line 
of the CuO$_{2}$ square lattice, to a collinear incommensurate one, 
which is modulated along the collinear line of the CuO$_{2}$ square lattice, 
systematic $\mu$SR studies~\cite{Ch.Niedermayer_98} have 
revealed that the spin freezing temperature, corresponding to the 
spin-glass transition, changes continuously across the boundary.
Thus, it is important to clarify whether the spin-glass state is
essential for the superconductivity in this system.

In the present study, focusing on these unresolved issues, 
we have performed systematic magnetic susceptibility measurements on single 
crystals with $x=0.03, 0.04$ and $0.05$ and have determined the magnetic phase 
diagram of the lightly doped region from both the bulk  
susceptibility and recent neutron scattering experiments.\cite{waki_neut} 
The contents of this paper are as follows: 
the theoretical background of the spin-glass features, including a scaling hypothesis 
of the spin-glass order parameter, is presented in Sec.~\ref{sec_sg}. 
Section ~\ref{sec_exp} describes sample preparation and experimental details.
The results of the magnetic susceptibility measurements and a revised magnetic 
phase diagram are presented in Sec.~\ref{results}.
In Sec.~\ref{sec_dis} we discuss interpretations for some of the remarkable 
features we find, 
combining the susceptibility with previous neutron scattering 
and $\mu$SR measurements.

\section{Features of the canonical spin-glass order parameter}
\label{sec_sg}

Determination of the spin-glass state is not an easy task, since 
the effects of randomness do not appear to be unique. Furthermore the 
effect of frustration by doped holes in the 
LSCO system should be very strong due to 
the strong super-exchange interaction in the 
antiferromagnetic lattice in each CuO$_{2}$ plane. If the formation 
of the spin-glass state is commonly visible in a certain hole 
concentration range, we can expect that a strong local singlet 
formation between spins of doped holes at oxygen sites and nearest 
neighbor Cu$^{2+}$ spins must be an essential ingredient 
of the HTSC mechanism in this material.
	
In this paper we first demonstrate that the magnetic system we treat 
is a quenched spin-glass system, with holes introduced  
by random substitution of La site with Sr cations. Note that 
another doping case by an insertion of excess oxygens is supposed 
to be an annealed system, where the doped oxygens are staged and 
also ordered in each staged layer at slow cooling stage.
\cite{Wells_96,Xiong_96}
	
In the quenched spin-glass system, magnetic susceptibility is 
postulated to result from the sum of a temperature independent 
residual susceptibility and a Curie type magnetic susceptibility:
\begin{equation}
\chi = \chi_{0} + \frac{C}{T}(1-q) \label{eq:orderp}
\end{equation}
where $\chi_{0}$ is temperature-independent susceptibility, $C$ 
is the Curie constant, and $q$ is the spin-glass order parameter. 
Thus, non-zero spin-glass order parameter $q$ gives 
rise to deviations of the magnetic susceptibility from simple 
Curie law behavior below the spin-glass transition temperature, $T_{g}$. 
Usually the thermal evolution of $\chi$ shows a distinct cusp at 
$T_{g}$ which is characterized as a typical spin-glass feature. 

More rigorously, the spin-glass order parameter should obey
the scaling relation~\cite{Malozemoff_83} described as
\begin{equation}
q = |t|^{\beta} \cdot f_{\pm}(H^2|t|^{-\beta-\gamma}) \label{eq:scaling}
\end{equation}
\begin{equation}
t = \frac{T-T_{g}}{T_{g}}. \label{eq:normalt}
\end{equation}
where $H$ is applied magnetic field, $\beta$ and $\gamma$ are 
critical exponents, and $T_{g}$ is spin-glass transition temperature.
The scaling functions $f_{+}$ and 
$f_{-}$ are defined for $T > T_{g}$ and $T < T_{g}$, respectively.
The critical exponents, 
$\beta$ and $\gamma$, should be compared with the values deduced by the renormalization 
group theory of $0.5 < \beta < 1$ and $\gamma = 3 \pm 1$. 
In fact, earlier measurements of Chou {\it et al.}~\cite{F.C.Chou_95} showed that 
the spin-glass state of the La$_{1.96}$Sr$_{0.04}$CuO$_{4}$ sample 
obeys the scaling relation with
$\beta \sim 0.9$ and $\gamma \sim 4.3$. These results stand as an important
reference for the present experiments.
(Based on their results, we reevaluated the hole concentration of the sample
used by Chou {\it et al.} and find a value somewhat lower than the quoted $x=0.04$;
see Section~\ref{results}.)

Another important piece of evidence for the spin-glass state is the remanent 
magnetization below $T_{g}$ when an external field is turned 
off after crossing $T_{g}$ (Field cooling effect). 
It shows the distinct difference of the magnetic susceptibilities 
between the zero field cooling and field cooling process below $T_{g}$. 
After the external field is switched off, the remanent magnetization typically 
relaxes following a stretched exponent function of time $\tau$.
\begin{equation}
M(\tau) = M(0)exp[\alpha \tau^{(1-n)}] \label{eq:remanent}
\end{equation}

Theoretical predictions give $1-n = 1/3$, which is often found 
in typical spin-glass compounds. In fact, we confirm that the 
stretching exponent for the magnetization memories of the present crystals have the same 
values as in the experiments reported in Ref.8.

\section{Sample preparation and experimental details}
\label{sec_exp}

Single crystals of La$_{2-x}$Sr$_{x}$CuO$_{4}$ with $x = 0.03, 0.04$ and $0.05$ 
were grown by TSFZ 
(Travelling-Solvent Floating-Zone) method using a standard 
floating-zone furnace with some improvements.\cite{lee_98}
Dried powders of La$_{2}$O$_{3}$, SrCO$_{3}$ and CuO of 99.99~\% 
purity were used as starting materials for feed rods and solvent.
The starting materials were mixed and baked in air at 850$^{\circ}$C, 
950$^{\circ}$C and 1000$^{\circ}$C for 24 hours each  
with a thorough grinding between each baking. 
After this process, we confirmed by X-ray powder diffraction that the 
obtained powder samples consisted only of a single 2-1-4 phase. 
Solvents with the composition
of La$_{2-x}$Sr$_{x}$CuO$_{4}$ : CuO = 35 : 65 in molar 
ratio were utilized in all growths. 
The growth conditions were basically 
the same as those used for the $x = 0.15$ crystal reported in Ref.19 
%\cite{lee_98}
except that excess CuO of $\sim 1$~mol\% was added into the feed rods to increase 
their packing density as well as to compensate the loss of 
CuO vaporizing in the melting process. 
The final 
%==================================================================
\begin{figure} 
 \centerline{\epsfxsize=3.2in\epsfbox{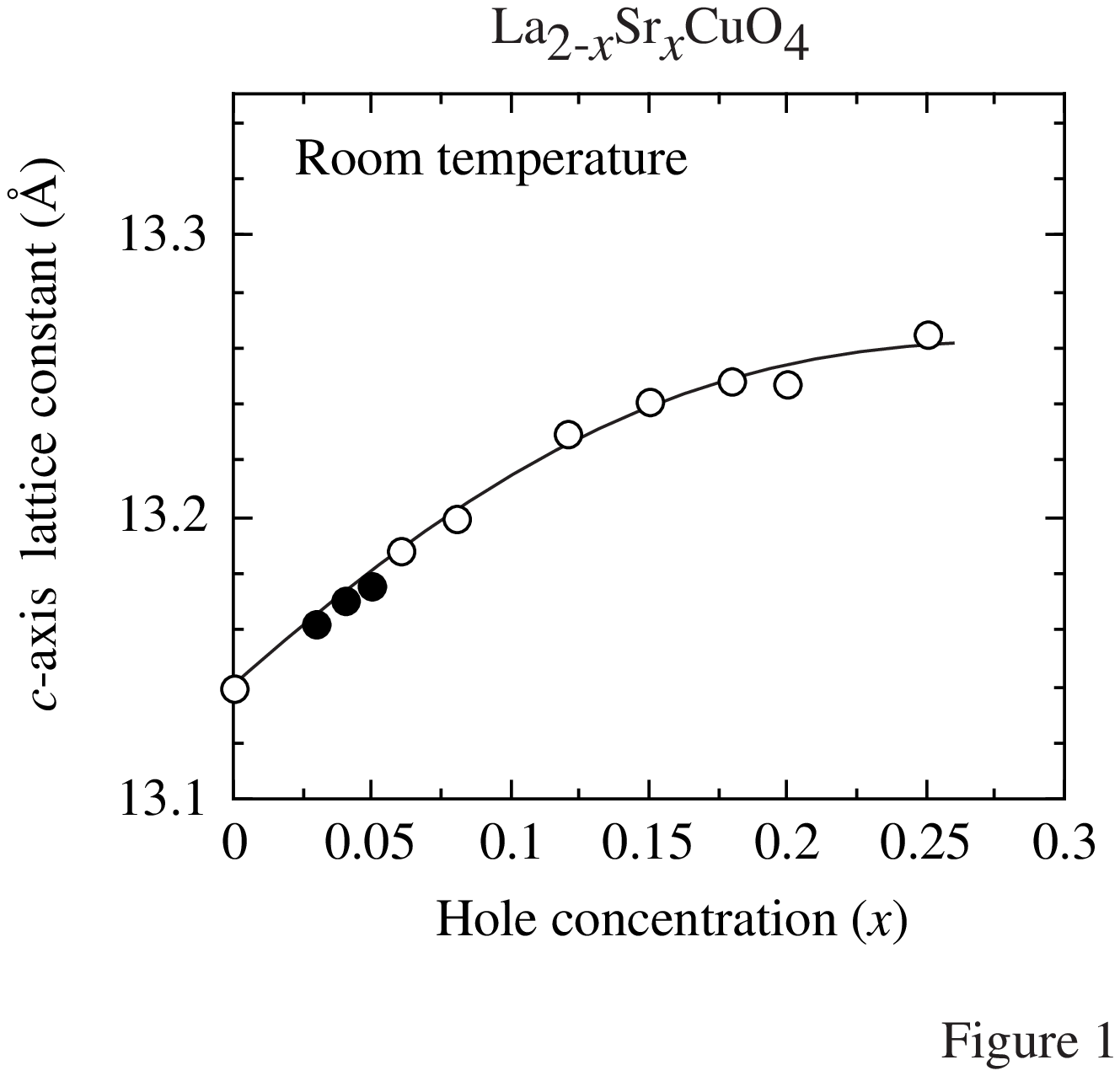}}
  \caption{Variation of the $c$-axis lattice constant versus  
  hole concentration at the room temperature. Closed circles 
  are data of the present crystals. Open circles, except for $x=0$, 
  are data from Yamada \protect{\it et al.}\protect
  \cite{K.Yamada_98} The value at $x=0$ was measured using a 
  single crystal annealed in Ar atmosphere. The solid 
  line is a guide to the eye.}
 \label{Fig:c-axis}
\end{figure}
%===================================================================
\noindent
crystals were typically 6~mm in diameter and 30~mm in length.

In order to characterize the Sr concentration of each crystal, 
the $c$-axis lattice constant was measured by X-ray powder diffraction. 
We found that lightly doped LSCO 
crystals tend to become oxygenated  
by the oxygen atmosphere of the melt-grown process; therefore the as-grown 
crystals 
were annealed in flowing Ar at 900$^{\circ}$C for 12 hours so as to 
reach the stoichiometric oxygen content before the X-ray diffraction
measurements. 
In fact, an iodometric titration analysis on the $x=0.03$ crystal revealed that 
the oxygen content of the Ar-annealed crystal is closer to the stoichiometric 
oxygen content than that of the as-grown crystal.
Figure~\ref{Fig:c-axis} shows the dependence of the $c$-axis lattice constant 
on hole concentration. 
The data for $x \geq 0.06$ (open circles) are from Ref.3, 
%\cite{K.Yamada_98} 
and the results for the present crystals are plotted by closed circles. 
In addition, data for $x=0$ was determined from a single crystal of 
La$_{2}$CuO$_{4}$ that was grown by the method reported in Ref.19 
%\cite{lee_98} 
and annealed in Ar atmosphere at 900$^{\circ}$C for 12 hours. 
As clearly shown in Fig.\ref{Fig:c-axis}, 
the $c$-axis lattice constants of the lower doped samples extrapolate smoothly from 
the published data.
This smooth extrapolation indicates that the hole concentration of our crystals 
correspond closely to the expected values of $x=0.03, 0.04$ and $0.05$. 
We note that our results are subtly lower than the powder sample data reported by 
Takayama-Muromachi {\it et al.}\cite{Takayama} in the region of $x \leq 0.06$. 
We speculate that this small difference is caused by an effect of the Ar annealing
performed on the present crystals. 

The magnetic susceptibility measurements were performed using a standard 
Quantum Design SQUID 
%========================================================================
\begin{figure}
 \centerline{\epsfxsize=3.2in\epsfbox{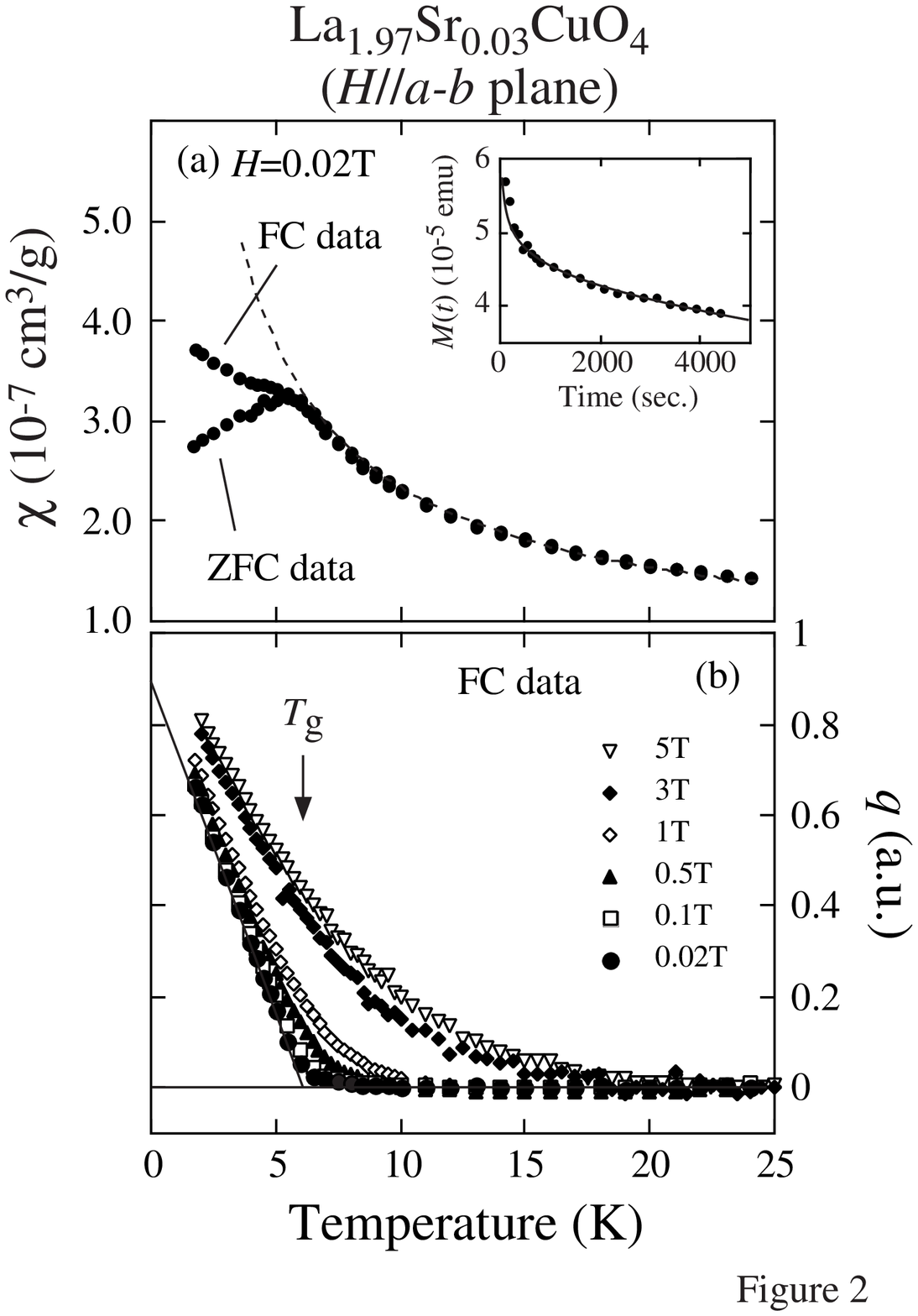}}
  \caption{(a) Temperature dependence of the in-plane magnetic 
  susceptibility for the $x=0.03$ sample. A magnetic field of 
  $0.02$ Tesla was applied parallel to the Cu0$_{2}$ planes.  
  Both ZFC and FC data are included. A dashed line indicates the 
  Curie law corresponding to the susceptibility with $q=0$ as determined 
  by a least-square fitting at higher temperatures
  $(10~k \leq T \leq 70~K)$. The inset shows a time dependence of the 
  remanent magnetization. The solid line is a least-square fit  
  to Eq.~\protect\ref{eq:remanent}. (b) Temperature dependence of spin-glass 
  order parameter $q$ determined from the FC data by Eq.~\protect\ref{eq:orderp}
  at several applied fields $(0.02~T 
  \leq H \leq 5~T)$. The solid line is a least-square fit of  
  the $H=0.02$~T data to the equation of $q \propto (T_{g}-T)^{\beta}$.}
 \label{Fig:in-plane}
\end{figure}
%=========================================================================
\noindent
magnetometer. 
Measurements were made on each crystal under various applied fields in the range  
$0.02~{\rm T} \leq H \leq 5$~T either parallel or perpendicular to 
the CuO$_{2}$ planes.
We did not specify the field direction within the plane.
Data 
was measured either by cooling with the field applied (FC) or 
by applying field after cooling in zero field (ZFC). 
Note that the CuO$_{2}$ 
plane corresponds to the $a-b$ 
plane of the orthorhombic 
$Bmab$ crystallographic notation, which is utilized throughout the paper.

\section{Magnetic susceptibility}
\label{results}

For all of the samples we studied, the magnetic susceptibility under 
a magnetic field parallel to the CuO$_{2}$ 
%====================================================================
\begin{figure}
 \centerline{\epsfxsize=3.2in\epsfbox{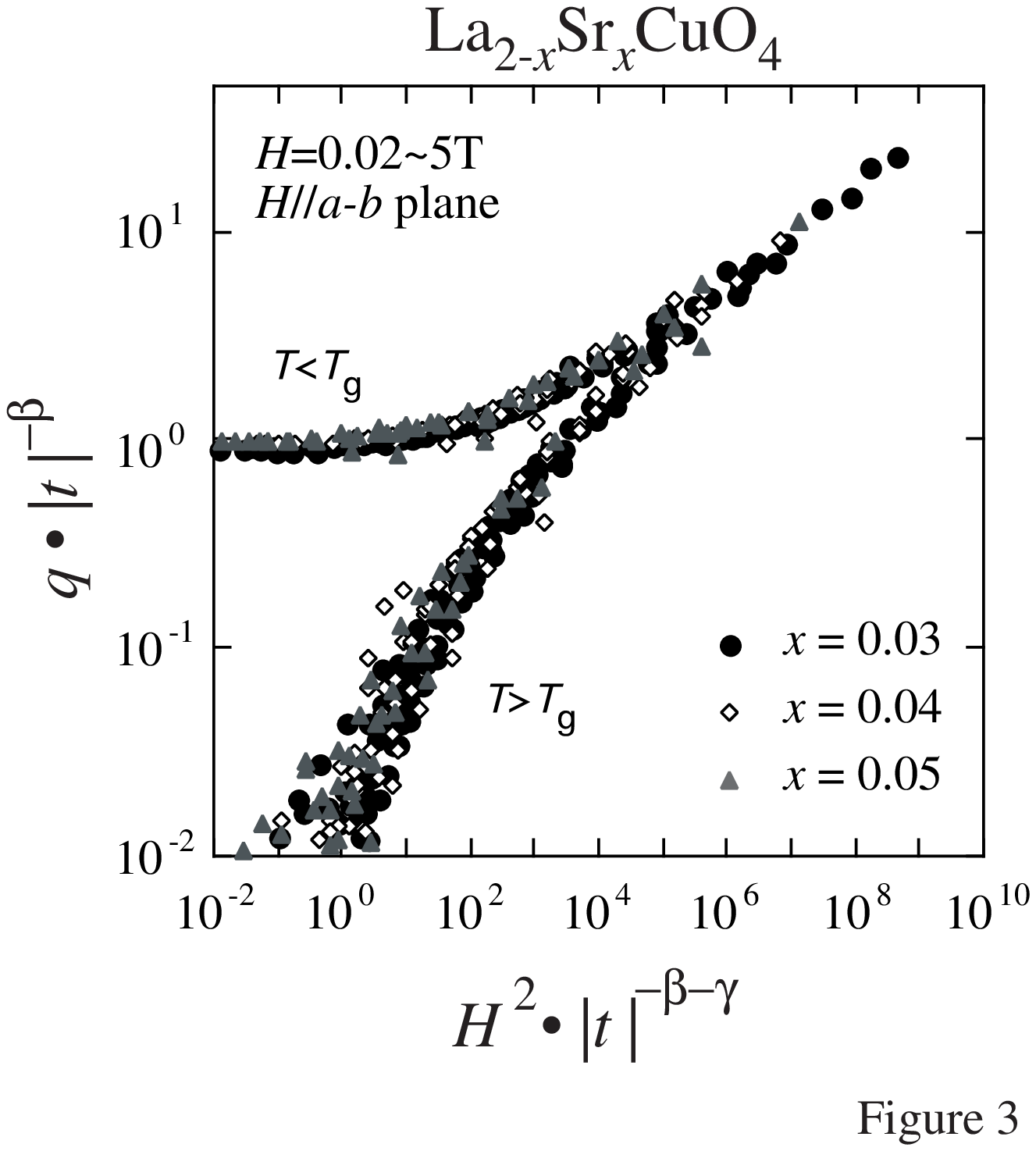}}
  \caption{Universal curves plotted for the scaling relation of 
  Eq.~\protect\ref{eq:scaling} using the in-plane FC data from the $x=0.03, 
  0.04$ and $0.05$ samples with applied fields of $0.02~T \leq H 
  \leq 5~T$. The spin-glass exponents $\beta$ and $\gamma$ are 
  $0.97(\pm0.05)$ and $3.2(\pm0.5)$, respectively.}
 \label{Fig:scaling}
\end{figure}
%====================================================================
\noindent
plane, which we call in-plane 
magnetic susceptibility, exhibits qualitatively the same temperature dependence.
As a typical example Fig.~\ref{Fig:in-plane}(a) shows the temperature 
dependence of the in-plane susceptibility of the $x=0.03$ sample.
A clear difference exists between the FC and 
ZFC data corresponding to hysteresis 
characteristic of spin-glasses.
The inset of Fig.~\ref{Fig:in-plane}(a) shows the time dependence of the remanent 
magnetization after turning off an applied field of 1~T.
This time dependence is well fit by Eq.~\ref{eq:remanent} with $1-n=1/3$ (solid line).
These facts provide a direct evidence that a canonical spin-glass state exists 
in the CuO$_{2}$ planes in this system.

Below the spin-glass transition temperature $T_{g}$  
the magnetic susceptibility starts to deviate from Curie law, as 
indicated by a dashed line in Fig.~\ref{Fig:in-plane}(a). 
The dashed line of the Curie law was calculated by a least-squares fit to 
the data between 10K and 70K. 
The spin-glass order parameter $q$ can be evaluated from the degree of the 
deviation using Eq.~\ref{eq:orderp}. 
Figure~\ref{Fig:in-plane}(b) shows the thermal variation of $q$ calculated from the FC data 
in various applied fields. 
$T_{g}$ was determined by a least-square fit to the temperature dependence of 
$q$ at $H=0.02~$T using the equation $q \sim (T_{g}-T)^{\beta}$. 
The solid line in Fig.~\ref{Fig:in-plane}(b) shows the result of the fit with  
$T_{g}=6.3(\pm 0.5)$~K and $\beta = 0.97(\pm 0.05)$. 
Similarly, $T_{g}$ for the $x=0.04$ and $0.05$ samples are determined to be $5.5(\pm 0.5)$~K 
and $5.0(\pm 0.5)$~K with the same value of $\beta$, respectively.

To characterize further the spin-glass properties in this system, we verify the 
scaling hypothesis described by Eq.~\ref{eq:scaling}. 
Figure~\ref{Fig:scaling} shows the scaling of the in-plane magnetic 
%=========================================================================
\begin{figure}
 \centerline{\epsfxsize=3.2in\epsfbox{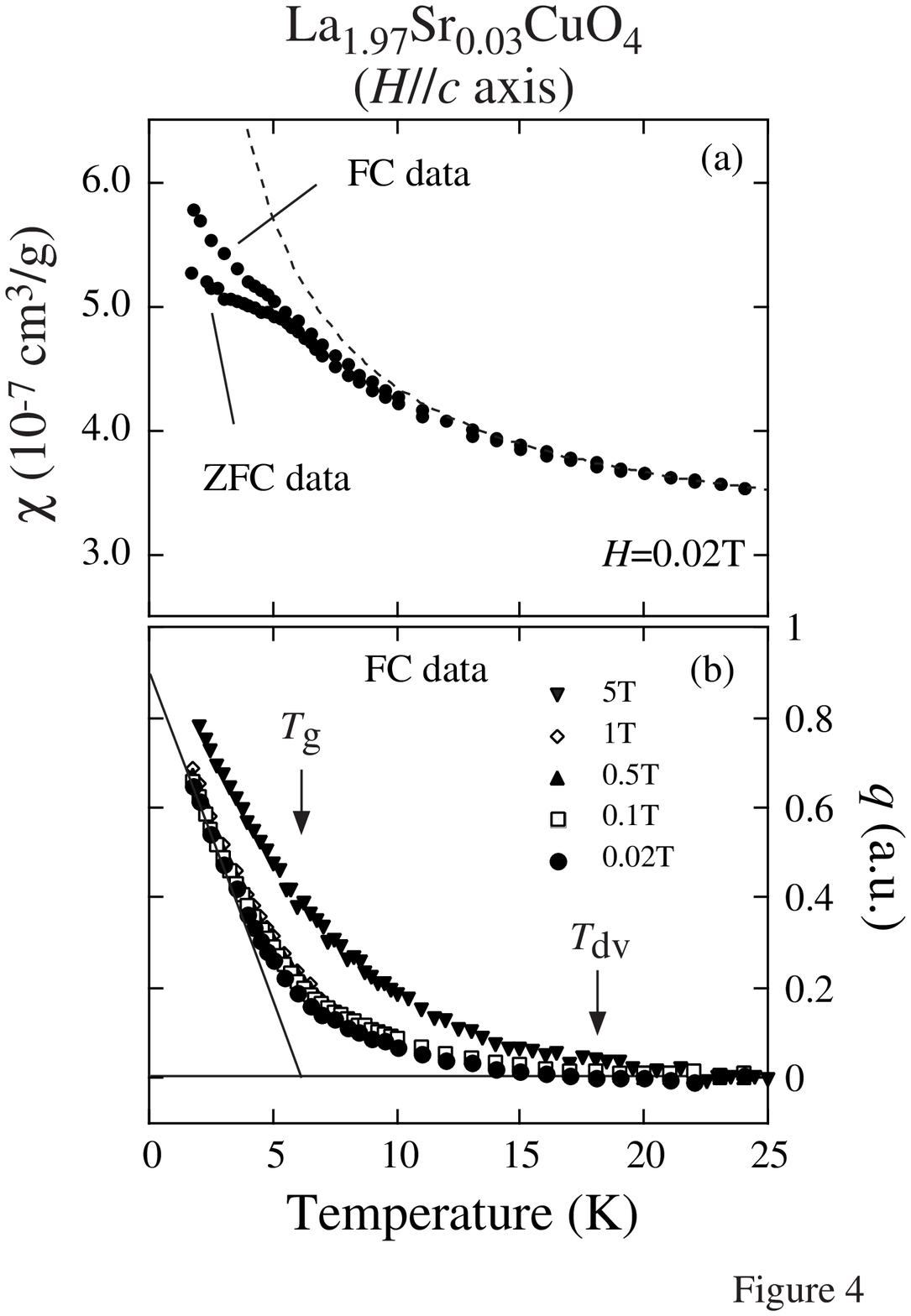}}
  \caption{(a) Temperature dependence of the out-of-plane magnetic 
  susceptibility of the $x=0.03$ sample. A magnetic field of $0.02$ 
  Tesla was applied perpendicular to the CuO$_{2}$ planes. The dashed 
  line shows the Curie law corresponding to a susceptibility with $q=0$. 
  (b) Temperature dependence of spin-glass order parameter $q$ calculated 
  from the FC data at several applied fields $(0.02~T \leq H \leq 5~T)$. The 
  solid line is a fit result of the in-plane data. The arrow indicates $T_{g}$ 
  determined by the in-plane data.}
 \label{Fig:out-of-plane}
\end{figure}
%===========================================================================
\noindent
susceptibility. 
The scaling relation is well satisfied for all samples with 
the same value of the critical exponents. 
This fact indicates that the spin-glass behavior in the 
in-plane susceptibility is common to the $x=0.03, 0.04$ 
and $0.05$ samples, and that these samples exhibit a canonical quenched spin-glass state. 
Furthermore, the critical exponents $\beta$ and $\gamma$ obtained from the universal 
plots are $0.97(\pm0.05)$ and $3.2(\pm0.5)$, which are consistent with those of 
typical canonical spin-glass materials.

In contrast, the magnetic susceptibility under a magnetic field along the out-of-plane
direction, which we call out-of-plane magnetic susceptibility, 
shows behavior different from typical spin-glasses. 
The temperature dependence of the out-of-plane susceptibility of the $x=0.03$ sample 
is shown in Fig.~\ref{Fig:out-of-plane}(a) as an example. 
A difference between FC and ZFC data is observable at low $T$, and 
a small shoulder appears at $T_{g}$ determined by the in-plane measurements. 
These features of the out-of-plane susceptibility are qualitatively similar 
to those of the in-plane 
%=========================================================================
\begin{figure}
 \centerline{\epsfxsize=3.2in\epsfbox{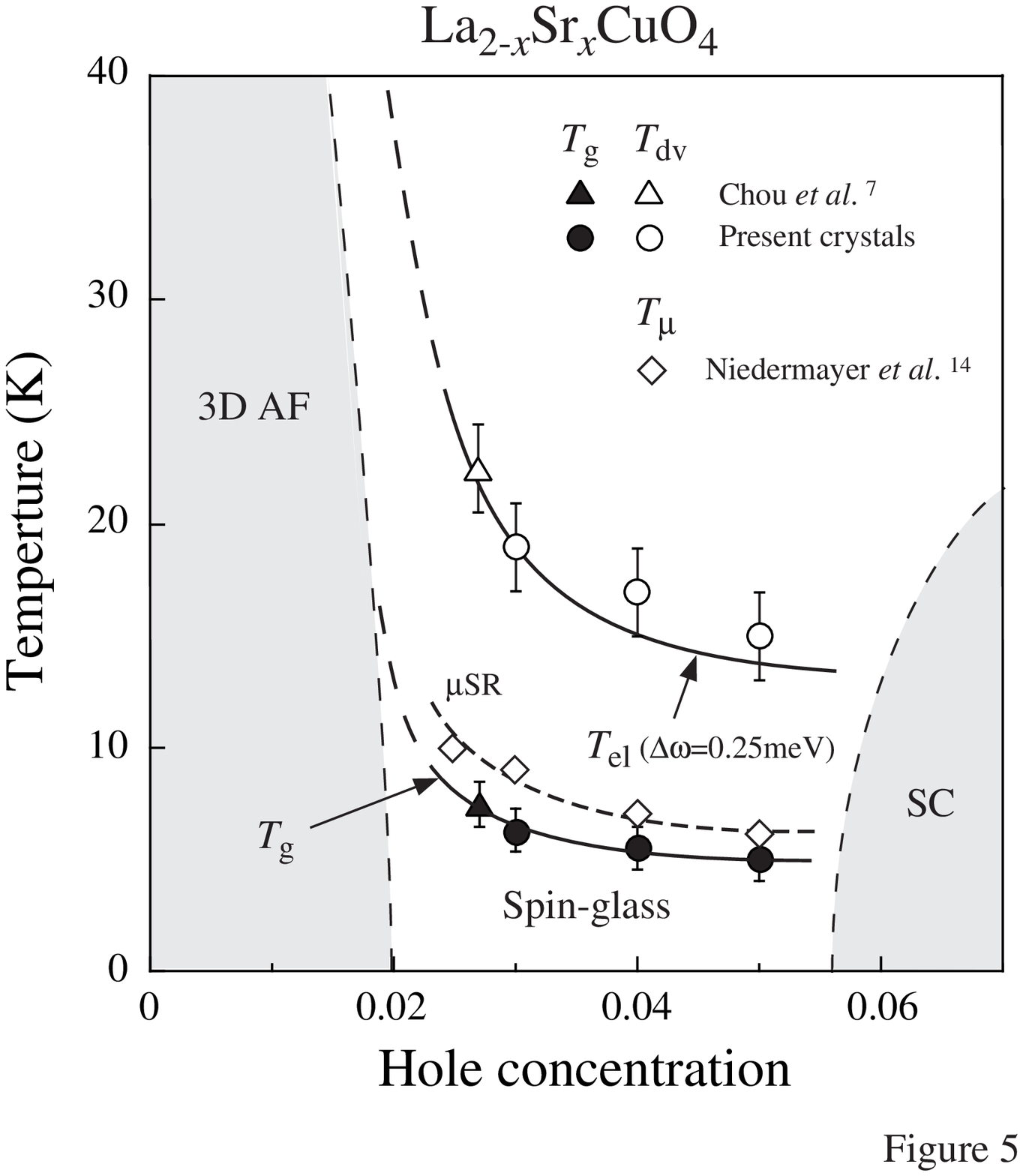}}
  \caption{Magnetic phase diagram for the lightly doped region. The circles are 
  the data of the present crystals and the triangles are those reported by Chou {\it et al.}
  \protect\cite{F.C.Chou_95} Closed and open symbols indicate $T_{g}$ and 
  $T_{dv}$, respectively, as determined by the magnetic susceptibility measurements. 
  The line for $T_{g}$ is a guide to the eye and the line for $T_{el}$ is from 
  Ref.11. Open diamonds indicates the temperatures 
  where magnetic signals are observed in the $\mu$SR 
  measurements.\protect\cite{Ch.Niedermayer_98}}
 \label{Fig:diagram}
\end{figure}
%==========================================================================
\noindent
spin-glass behavior. 
However, a remarkable difference is 
that the out-of-plane susceptibility 
deviates from a simple Curie law at a temperature well above $T_{g}$, 
as clearly 
shown in Fig.~\ref{Fig:out-of-plane}(a).
Such a deviation is observed in all the samples and was also reported by 
Chou {\it et al.}~\cite{F.C.Chou_95} 
This deviation indicates that, under the same analysis as for the in-plane 
susceptibility, $q$ increases from zero far above $T_{g}$. 
This feature is more clearly visible in a calculation of $q$ from the FC data,
as shown in Fig.~\ref{Fig:out-of-plane}(b). 
The onset temperatures for these deviations, $T_{dv}$, are found to be 
$19(\pm 2)$~K, $17(\pm 2)$~K and $15(\pm 2)$~K for $x=0.03, 0.04$ and $0.05$, respectively. 
In addition, the spin-glass order parameter of the out-of-plane magnetic susceptibility 
does not obey the scaling relation of Eq.~\ref{eq:scaling}. 
This result suggests that another magnetic mechanism rather than spin-glass behavior 
drives the deviation of the out-of-plane susceptibility from the simple Curie 
type paramagnetic behavior. 
 
We summarize the magnetic susceptibility data in Fig.~\ref{Fig:diagram}
with a magnetic phase diagram that includes results from 
Chou {\it et al.}~\cite{F.C.Chou_95} on an $x=0.04$ crystal 
and $\mu$SR results reported by Niedermayer {\it et al.}~\cite{Ch.Niedermayer_98}
The closed and open circles indicate $T_{g}$ and $T_{dv}$ of 
the present crystals, respectively, while the closed and open triangles are those of the 
``$x=0.04$" crystal of Chou {\it et al.}
Previous report on a sample of 
La$_{2-x}$Bi$_{x}$CuO$_{4+\delta}$ whose hole concentration is just 
above the boundary of 3D AF state showed a spin-glass transition with 
$T_{g} \simeq 17$~K~\cite{waki_97};
therefore, we draw an $x$-dependence line of $T_{g}$ as a guide to the eye which 
reaches 17~K at $x=0.02$. 
From this line, the actual hole concentration of the ``$x=0.04$" sample of 
Chou {\it et al.} can be estimated to be 0.027.\cite{estimation} 
Thus we treat the data of Chou {\it et al.} as $x=0.027$ in the present paper. 

According to the recent neutron scattering experiments reported by 
Wakimoto {\it et al.}~\cite{waki_neut,waki_full} using the same crystals as 
those in the present experiment, elastic incommensurate magnetic peaks exist  
around the $(\pi, \pi)$ position in reciprocal space. 
The solid line of $T_{el}$($\Delta\omega = 0.25$~meV) in Fig.~\ref{Fig:diagram} 
shows the $x$-dependence of the onset temperature where the elastic 
magnetic peaks become observable with the energy resolution of $\Delta\omega=0.25$~meV. 
$T_{dv}$ exhibits qualitatively the same $x$-dependence as 
that of $T_{el}$($\Delta\omega = 0.25$~meV). 
The physical meaning of this feature is discussed in the next section.

The temperatures below which a magnetic signal is observed in the $\mu$SR 
measurements of Niedermayer {\it et al.}~\cite{Ch.Niedermayer_98} are also
plotted as open diamonds in Fig.~\ref{Fig:diagram}. (Hereafter, we label
these temperatures as $T_{\mu}$.) 
Niedermayer {\it et al.} equated $T_{\mu}$ with the spin-glass transition temperature. 
However, in our phase diagram, there is a small discrepancy between $T_{\mu}$ and $T_{g}$. 
%**********ADDITIONAL PART**********
Our $T_{g}$ values for $x=0.03$ and $0.04$ are very close to the spin 
glass freezing temperature determined by the $^{139}$La NQR measurements 
in Ref.13.
%***********************************
We believe that the differences among $T_{el}$, $T_{\mu}$ and $T_{g}$ arise from 
the differences in the observation time scales for the different experimental methods.
In fact, Keimer {\it et al.}~\cite{B.Keimer_92} reported that $T_{el}$ depends on
the instrumental energy resolution, which varies inversely with observation time scale. 
Furthermore, the nature of the magnetic susceptibility measured by SQUID is perfectly static. 
These facts indicate that the magnetic correlations in this system are essentially 
quasi-static, and, specifically, that with decreasing temperature the spin system freezes into
a cluster spin-glass state consisting of domains in which spins are 
antiferromagnetically correlated.

\section{Discussion}
\label{sec_dis}

The present results of the magnetic susceptibility measurements have revealed 
the existence of a common canonical spin-glass state for La$_{2-x}$Sr$_{x}$CuO$_{4}$
in the insulating region of $0.03 \leq x \leq 0.05$. 
In this section, we discuss the nature of the spin correlations specifically
to reconcile this finding with the recent 
neutron scattering results on samples in the same hole concentration range. 

First, we discuss the $x$-dependence of the characteristic temperatures
$T_{g}$, $T_{el}$ and $T_{dv}$.
In Fig.~\ref{Fig:diagram}, all of these temperatures exhibit qualitatively 
similar $x$-dependences.
Since the difference between $T_{g}$ and $T_{el}$ reflects the different 
time scales of the experimental probes in observing the freezing spin system, 
the similarity between the $x$-dependences of $T_{g}$ and $T_{el}$ indicates 
that the freezing process into the cluster glass state is similar 
for the 
%======================================================================
\begin{figure}
 \centerline{\epsfxsize=3.1in\epsfbox{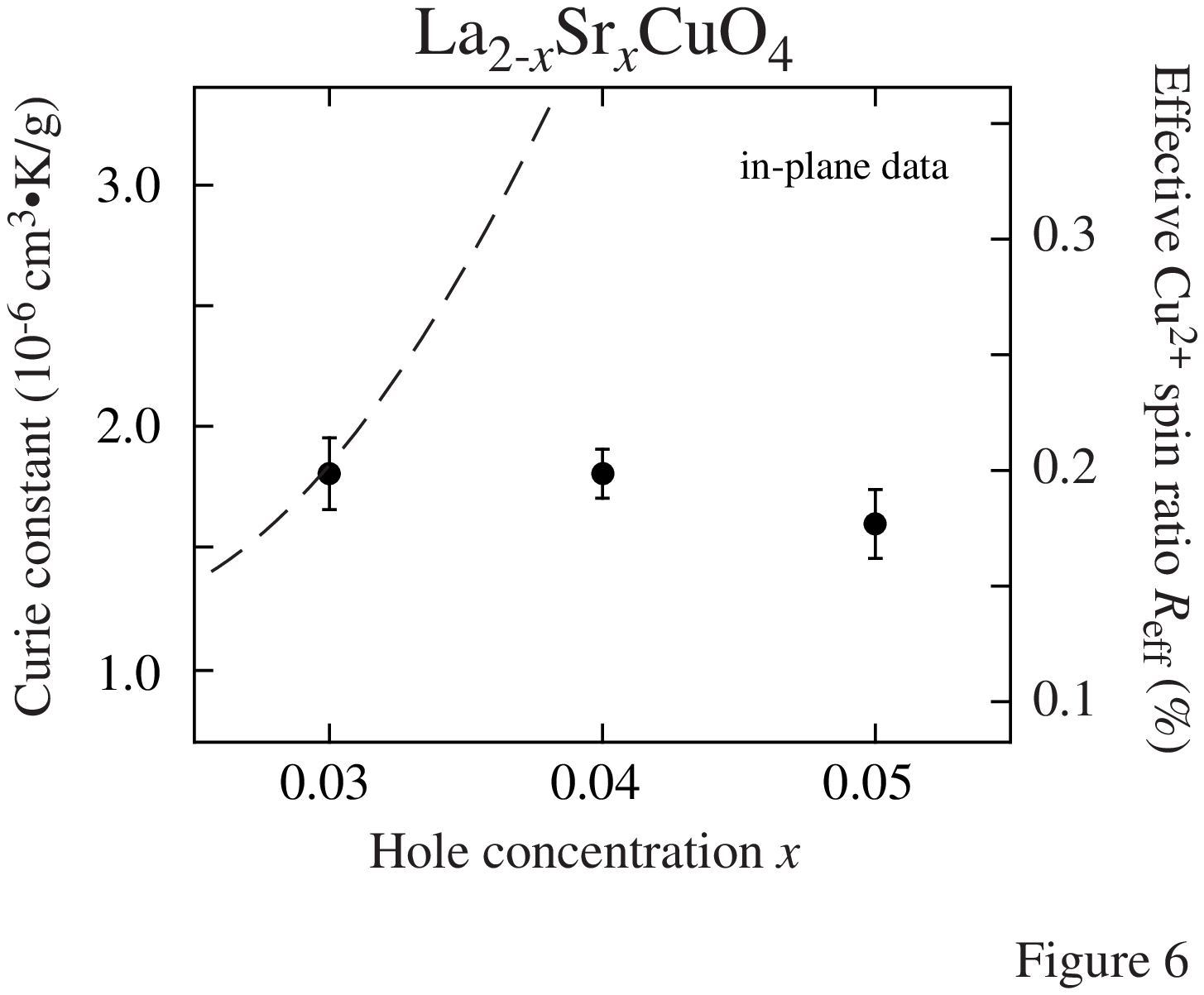}}
  \caption{Hole concentration dependence of the Curie constant $C$ 
  determined by the least-square fit to the in-plane magnetic susceptibility 
  data. The left vertical axis shows the Curie constant while the right 
  vertical axis shows the effective Cu$^{2+}$ spin ratio $R_{eff}$. A 
  dashed line corresponds to the theoretical prediction by Gooding {\it et 
  al.}~\protect\cite{Gooding} (see text.)}
 \label{Fig:curie}
\end{figure}
%========================================================================
\noindent
samples in this hole concentration range.
Although the $T_{dv}$ values sit on the 
$T_{el}$($\Delta\omega = 0.25$~meV) line,
the significant feature is not their equivalent values but their qualitatively 
similar $x$-dependence.
As mentioned in Section \ref{results}, the position of $T_{el}$ 
is arbitrary to the extent that it depends on 
the instrumental energy resolution width $\Delta\omega$.
The similar $x$-dependences of $T_{dv}$ and $T_{el}$ suggest a correlation 
between the formation of the spin clusters 
and the deviation of the out-of-plane susceptibility from 
simple Curie type behavior.
There may also be a relation between this deviation from the Curie law 
and the development of short magnetic correlations along the out-of-plane direction 
as observed by neutron scattering measurements for 
$x=0.03$ and $0.05$ at low temperature.~\cite{waki_neut,waki_full}

Next, we discuss the spin-glass properties of the in-plane magnetic susceptibility,
particularly,  
the $x$-dependence of the Curie constant shown in 
Fig.~\ref{Fig:curie}. 
The Curie constants, calculated from a least-square 
fit to the in-plane susceptibility data between 10K and 70K,  
are essentially independent of $x$ for this hole concentration range.
The right vertical axis of Fig.~\ref{Fig:curie} indicates the scale of the 
ratio $R_{eff}=N_{eff}/N_{all}$, 
where $N_{eff}$ is the number of effective Cu$^{2+}$ spins giving the simple Curie 
type paramagnetic susceptibility and 
$N_{all}$ is the total number of Cu$^{2+}$ spins per unit volume. 
$N_{eff}$ is related to the Curie constant by the following formula:
\begin{equation}
C=N_{eff} \frac{(g \mu_{\rm B})^{2} S(S+1)}{3k_{\rm B}} \label{eq:curieconst}
\end{equation}
where $g=2$, $S=1/2$ and $k_{B}$ is Boltzmann's constant. 

As pointed out by Gooding {\it et al.},~\cite{Gooding}
if the cluster spin-glass state is realized in this system, 
each cluster consisting of an odd number of spins behaves like a single spin 
yielding a Curie type paramagnetic behavior. 
Hence, the number of effective 
Cu$^{2+}$ spins should be half of the total number of the clusters. 
Therefore, $R_{eff}$ can be described using an average cluster size $L$ 
(expressed in the unit of the nearest neighbor Cu-Cu distance) as 
\begin{equation}
R_{eff}=\frac{1}{2L^{2}}.  \label{eq:prediction}
\end{equation}
From a numerical calculation, 
Gooding {\it et al.}~\cite{Gooding} reported that $L$ is given by 
$L=Ax^{-\eta}$ ($A \sim 0.49$ and $\eta \sim 0.98$) on the basis of 
the cluster spin-glass model with a random distribution of doped holes.
In this model, the doped holes form the boundaries between the clusters.
Their result is indicated by a dashed line in Fig.~\ref{Fig:curie}.
Although the $R_{eff}$ value for $x=0.03$ is close to the prediction of Gooding 
{\it et al.}, the $R_{eff}$ values are independent of $x$, with those for $x=0.04$ and 
$0.05$ significantly smaller than the dashed line.
The constant value of $R_{eff}$ combined with Eq.~\ref{eq:prediction}, means that 
the cluster size is independent of the hole concentration, 
instead of decreasing with increasing doping level
as predicted for a random distribution of holes 
located on the cluster boundary. 
 
A possible interpretation for these features is that the doped holes are
distributed also inside the clusters as well as on the boundaries.
Recent neutron scattering results~\cite{waki_full,Matsuda_incomme} 
demonstrate that quasi-static incommensurate spin correlations exist in the spin-glass region.
This may imply that the holes inside the clusters form charge stripes and 
each cluster includes anti-phase antiferromagnetic domains divided by these charge stripes.
In this model, the $x$-independence of the cluster size suggests that 
only the distance between nearest neighbor charge stripes 
(i.e. 1/incommensurability) within the clusters varies with the hole concentration.
Consistent with this picture, the incommensurability of the quasi-elastic peaks 
observed by neutron scattering experiments~\cite{waki_full,Matsuda_incomme} increases linearly 
with the hole concentration $x$ in the spin-glass region where we observe 
no change in the Curie constant.
The mechanism of cluster boundary formation is still an open question.
Possible constituents of the cluster boundary are, for example, a small amount of holes 
distributed outside the charge stripes or displacements of the stripes.
A correct model for the cluster glass state including charge stripes needs further 
clarification.

Briefly, we should note another possibility for the discrepancy between 
the theoretical line and the present results for $R_{eff}$.
In the model of Gooding {\it et al.}, the spin-glass cluster size is supposed to be 
equivalent to the instantaneous magnetic correlation length.
However, the $R_{eff}$ values correspond to a time-averaged 
cluster size rather than instantaneous one 
since the $R_{eff}$ values are obtained by magnetic susceptibility measurements.
Additional careful investigations are required to distinguish the static from 
dynamic magnetic properties. 

To conclude, the present results combined with recent neutron scattering and 
$\mu$SR measurements indicate that the spin system for La$_{2-x}$Sr$_{x}$CuO$_{4}$ 
in the range $0.03 \leq x \leq 0.05$
freezes into a cluster spin-glass state at low temperatures while maintaining  
antiferromagnetically correlated spins in each cluster. 
Inside the clusters, it is likely that the doped holes are inhomogeneously 
distributed. A stripe structure is suggested by  
the elastic incommensurate peaks observed in the neutron 
scattering.~\cite{waki_full,Matsuda_incomme}

One of the remaining important issues is the role of the cluster spin-glass state 
with respect to the superconductivity in the La$_{2-x}$Sr$_{x}$CuO$_{4}$ system.
Since the $x$-dependences of all the characteristic temperatures 
$T_{g}$, $T_{el}$ and $T_{dv}$ near the 
superconducting boundary are very small
in Fig.~\ref{Fig:diagram}, the spin-glass region may 
extend into the superconducting region as suggested by Niedermayer 
{\it et al.}~\cite{Ch.Niedermayer_98}.
In fact, neutron scattering reveals quasi-static incommensurate peaks 
in under-doped superconducting samples which may relate the 
spin-glass phase within the superconducting state.~\cite{waki0.06}
However, the spatial spin modulation of the spin-glass state in the 
superconducting phase should differ 
from that of the spin-glass phase in the non-superconducting state 
as demonstrated by recent neutron scattering.~\cite{waki0.06}
In order to clarify the role of the cluster spin-glass state in the HTSC mechanism,
further investigations are necessary of the spin-glass
features in the superconducting region compared with those  
in the insulating region.

\section{Acknowledgments}

We gratefully thank R.J. Birgeneau, F. C. Chou, K. Hirota, Y.-J. Kim, 
Y.S. Lee, R. Leheny, 
S. Maekawa, G. Shirane and S.M. Shapiro for 
invaluable discussions. 
We also thank M. Onodera for his technical assistance. 
The present work has been supported by a Grant-in-Aid for Scientific 
Research from Japanese Ministry of Education, Science, Sports and Culture, 
by a Grant for the Promotion of Science from the Science and Technology Agency, 
and by the Core Research for Evolutional Science and Technology (CREST). 

\vspace{5mm}
\noindent
* Present address: Massachusetts
Institute of Technology, 77 Massachusetts Ave., Cambridge, MA 02139, USA.


\begin{references}
\bibitem{estimation} This reestimation was also confirmed by a $x$-dependence of 
$T_{el}$ 
determined by the neutron scattering experiments in Ref.11.
\end{references}
\end{document}